\documentclass{aa} 
\usepackage{color}
\usepackage{comment}
\usepackage{natbib}
\usepackage{url}
\usepackage{hyperref}
\usepackage{subfigure}
\usepackage{xspace}
\usepackage[percent]{overpic}
\usepackage{subfloat}
\usepackage{lineno}

\newcommand{\nustar}{{\it NuSTAR}\xspace}

\newcommand{\src}{RX~J0440}
\graphicspath{{./}{figures/}}

\begin{document}

\title{Discovery of spin-phase-dependent QPOs in the supercritical accretion regime from the X-ray pulsar RX~J0440.9+4431}

\titlerunning{Appearance of triple transient QPOs (ATTO) in the X-ray pulsar RX~J0440.9+4431}
\authorrunning{Malacaria et al.}


\author{
Christian~Malacaria \inst{\ref{in:ISSI}}
\and Daniela~Huppenkothen \inst{\ref{in:Netherlands},\ref{in:api}} 
\and Oliver~J.~Roberts \inst{\ref{in:USRA}}
\and Lorenzo~Ducci \inst{\ref{in:IAAT}}
\and Enrico~Bozzo \inst{\ref{in:Geneva}}
\and Peter~Jenke \inst{\ref{in:CSPAR}}
\and Colleen~A.~Wilson-Hodge \inst{\ref{in:MSFC}}
\and Maurizio~Falanga \inst{\ref{in:ISSI}}
}
\institute{
International Space Science Institute, Hallerstrasse 6, 3012 Bern, Switzerland \label{in:ISSI} \\ \email{cmalacaria.astro@gmail.com}
\and SRON Netherlands Institute for Space Research, Niels Bohrlaan 4, 2333CA Leiden, The Netherlands \label{in:Netherlands}
\and Anton Pannekoek Institute for Astronomy, University of Amsterdam, Science Park 904, 1098 XH, Amsterdam, The Netherlands \label{in:api}
\and Science and Technology Institute, Universities Space and Research Association, 320 Sparkman Drive, Huntsville, AL 35805, USA \label{in:USRA}
\and Institut f\"ur Astronomie und Astrophysik, Kepler Center for Astro and Particle Physics, Universit\"at T\"ubingen, Sand 1, 72076 T\"ubingen, Germany \label{in:IAAT}
\and Department of Astronomy, University of Geneva, Chemin d'Ecogia 16, CH-1290 Versoix, Switzerland \label{in:Geneva}
\and University of Alabama in Huntsville (UAH), Center for Space Plasma and Aeronomic Research (CSPAR), 301 Sparkman Drive, Huntsville, Alabama 35899 \label{in:CSPAR}
\and ST 12 Astrophysics Branch, NASA Marshall Space Flight Center, Huntsville, AL 35812, USA \label{in:MSFC}
}

\abstract
{RX~J0440.9+4431 is an accreting X-ray pulsar (XRP) that remained relatively unexplored until recently, when major X-ray outburst activity enabled more in-depth studies. Here, we report on the discovery of ${\sim}0.2$ Hz quasi-periodic oscillations (QPOs) from this source observed with {\it Fermi}-GBM. The appearance of QPOs in RX~J0440.9+4431 is triple transient, that is, QPOs appear only above a certain luminosity, only at certain pulse phases (namely corresponding to the peak of its sine-like pulse profile), and only for a few oscillations at time.
We argue that this newly discovered phenomenon (with the appearance of triple transient QPOs -- or ATTO) 
occurs if QPOs are fed through an accretion disk whose inner region viscosity is unstable when exposed to mass accretion rate and temperature variations.
Such variations are triggered when the source switches to the supercritical accretion regime and the emission pattern changes. We also argue that the emission region configuration is likely responsible for the observed QPOs spin-phase dependence.
}

\keywords{magnetic fields -- pulsars: individual: RX~J0440.9+4431 --  stars: neutron -- X-rays: binaries}

\maketitle

\section{Introduction}\label{sec:introduction}

The source RX~J0440.9+4431 (\src\ hereafter) is an accreting X-ray pulsar (XRP) first discovered with \textit{ROSAT} \citep{Motch1997}, and showing X-ray pulsations approximately every $202\,$s \citep{Reig1999}.
The system is a Be/X-ray Binary with a B0 optical companion, LS V +44~17 \citep{Reig2005LSV}.
The source Gaia distance is $2.4\pm0.1\,$kpc in DR3 \citep{Gaia2022}.

After more than ten years of quiescence, \src\ exhibited a major outburst starting on December 2022 and lasting until April 2023 \citep{Nakajima2022, Mandal2023, Coley2023, Pal2023}.
Several observations have been dedicated to this bright event.
Multiple pieces of evidence show that during the latest outburst the source transitioned from the subcritical to the supercritical accretion regime \citep{Salganik2023, Doroshenko2023_LSV}.
These two regimes are characterized by deceleration of the accretion flow dominated at the neutron star (NS) surface by Coulomb interactions and above the NS surface by a radiative shock, respectively, and are separated by the so-called critical luminosity, L$_c$, on the order of $10^{37}\,$erg/s \citep{Basko+Sunyaev76, Becker+12, Mushtukov15_crit_lum}.

The source spectrum 
can be described with two Comptonized components both at low and high luminosity \citep{Tsygankov2012, Salganik2023}.
A double-hump spectrum was, up to now, observed (see, e.g., \citealt[and references therein]{Tsygankov2019b}) and theoretically motivated \citep{Mushtukov2021,Sokolova-Lapa2021} only at low accretion rates.
A tentative cyclotron resonant scattering feature (or cyclotron line) detection was reported by \citet{Tsygankov2012} at 32 keV 
which, however, has not been confirmed in other works. 
\src's pulse profile is characterized by a single-peak, sine-like shape, plus a narrow dip observed in the soft X-ray band \citep{Usui12} attributed to absorption from the accretion stream \citep{Galloway2001}.

In this work, we report on the discovery of~\mbox{transient} quasi-periodic oscillations (QPOs) observed by \textit{Fermi}-GBM in \src\ during its latest outburst in 2023.
The observed QPOs are triple transient as they appear only above the critical luminosity, only at certain pulse phases, and only for a few oscillations at time.
We propose that~the appearence of triple-transient quasi-periodic oscil\-lations -- or ATTO -- is linked to the beaming pattern in~the supercritical accretion regime coupled with the system geometry.

\section{Observations and data reduction}\label{sec:data_reduction}

\textit{Fermi}-GBM (GBM hereafter) is the lower energy instrument aboard the \textit{Fermi} Gamma-ray Space Telescope. It is an all-sky monitor instrument consisting of twelve sodium iodide (NaI) detectors and two bismuth germanate (BGO) detectors \citep{Meegan2009}.
GBM's configuration allows wide-field (>8 sr) broadband (8 keV - 40 MeV) observations of transient and persistent X-ray and $\gamma$-ray sources.
Although designed to hunt for gamma-ray bursts, thanks to its continuous timing and spectral characterization of transient sources GBM has proved excellent in studying accreting XRPs as well through the GBM accreting pulsars program (GAPP, see \citealt{Malacaria20}).

\begin{figure}
\centering 
\includegraphics[width=0.48\textwidth]{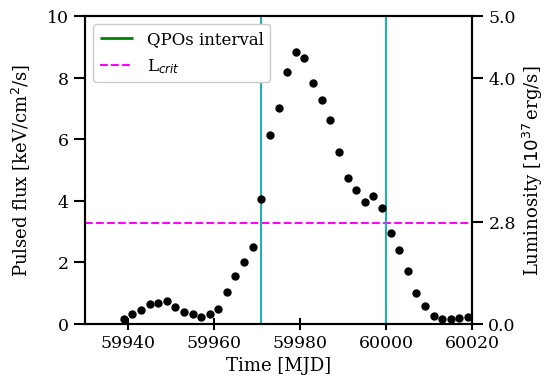}
\caption{GBM observations of \src\ during 2023 outburst. Black dots (with gray error bars) show the GBM (12--50 keV) three-day average pulsed flux.
The horizontal magenta dashed line indicates the critical luminosity at $2.8\times10^{37}\,$erg/s \citep{Salganik2023} as related to the right y-axis units. 
The nonlinearity in the right y-axis is due to a variable conversion factor at different outburst stages (see text).
The light green vertical continuous lines mark the interval when the source is showing QPO episodes.
\label{fig:outburst}}
\end{figure}

\textit{Fermi}-GBM triggered on \src\ several times during the brightest part of the outburst, from January 26 to February 25, 2023  (MJD 59970 -- 60000). 
Fig.~\ref{fig:outburst} shows \src\ as observed by the GAPP during the outburst\footnote{\url{https://gammaray.nsstc.nasa.gov/gbm/science/pulsars/lightcurves/rxj0440.html}}.
The luminosity scale shown in Fig.~\ref{fig:outburst} is obtained by converting the GBM pulsed flux to the \nustar\ luminosity values derived by \citet{Salganik2023}, with a conversion factor that depends on the outburst stage.
For the data analysis in this work, we considered time-tagged event (TTE -- 128 energy channels, 2 $\mu$s time binning) data from \src\ for events that triggered GBM on board.
All GBM triggers from \src\ occur at the peak of the sine-like pulse profile, and show one or more flares.
For each trigger we only analyze data from the three NaI detectors where the angle to the source is less than 60 deg.
Moreover, for our analysis we only considered triggers showing three or more flares (see Fig.~\ref{fig:230204}). 
A log of the analyzed triggers is shown in Table~\ref{tab:GBMtriggers}.
For data filtering and analysis we used the GBM Data 
Tools\footnote{\url{https://fermi.gsfc.nasa.gov/ssc/data/analysis/rmfit/gbm_data_tools/gdt-docs/}.} software, v1.1.1 \citep{Goldstein2022}.

\section{Data analysis and results}\label{sec:qpo_characterization}

We stacked TTE data for each trigger in Table~\ref{tab:GBMtriggers} and relevant detectors to accumulate statistics (the GBM trigger catalog is published online\footnote{\url{https://heasarc.gsfc.nasa.gov/W3Browse/fermi/fermigtrig.html}}).
Then, we limited the time window around the trigger time to one spin period (205 s) and used only data in the $8-70\,$keV energy band, above which the background dominates the signal. We used the combined event lists to construct light curves with a time resolution of $\Delta t = 0.00390625\;\mathrm{s}$, corresponding to a Nyquist frequency of $\nu_\mathrm{Ny} = 128 \; \mathrm{Hz}$.
We produced a periodogram over the whole pulse period using the fractional rms normalization \citep{belloni1990,miyamoto1992} for each light curve using the software package \textit{stingray} \citep{2019ApJ...881...39H, Huppenkothen2019} and averaged all considered periodograms to improve statistics. 
The resulting averaged periodogram is shown in Fig.~\ref{fig:psd_comp}.

\begin{figure} 
\includegraphics[width=.49\textwidth]{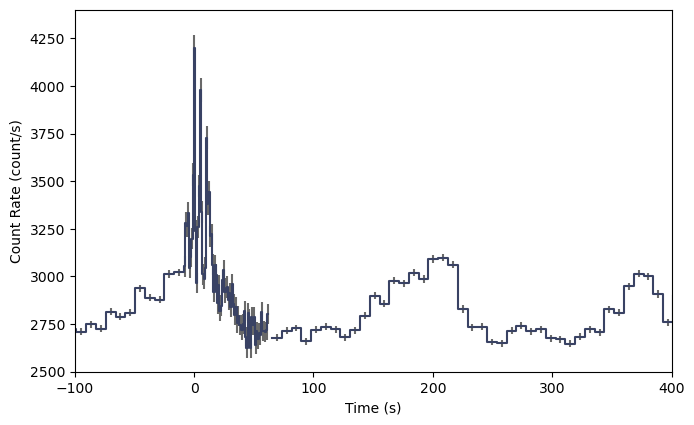}
\caption{GBM light curve of \src\ around trigger 230204.473 obtained by merging events from the NaI detectors N6, N7, and N8 in the 8--70 keV energy band and with a 1 s binning around the trigger (8 s otherwise). The sine-like long-term trend shows the source pulsation with an ${\sim}205\,$s periodicity. QPOs are evident at the peak of first pulse profile, around the GBM trigger time T0.
\label{fig:230204}}
\end{figure}
\begin{table}[!t]
\caption{Log of {\it {Fermi}}-GBM observations used in this work.  \label{tab:GBMtriggers}}
\centering
\begin{tabular}{cc}
\hline\hline 
Trigcat & Date (MJD) \\ 
\hline
 230203.275 & 59978.27\\
 230204.473 & 59979.47 \\
 230206.376 & 59981.38  \\
 230206.723 & 59981.72  \\
 230206.797 & 59981.79  \\
 230209.271 & 59984.27  \\
 230211.733 & 59986.73  \\
 230218.657 & 59993.66  \\
\hline
\end{tabular}
\end{table}

\begin{figure*}
\includegraphics[width=\textwidth]{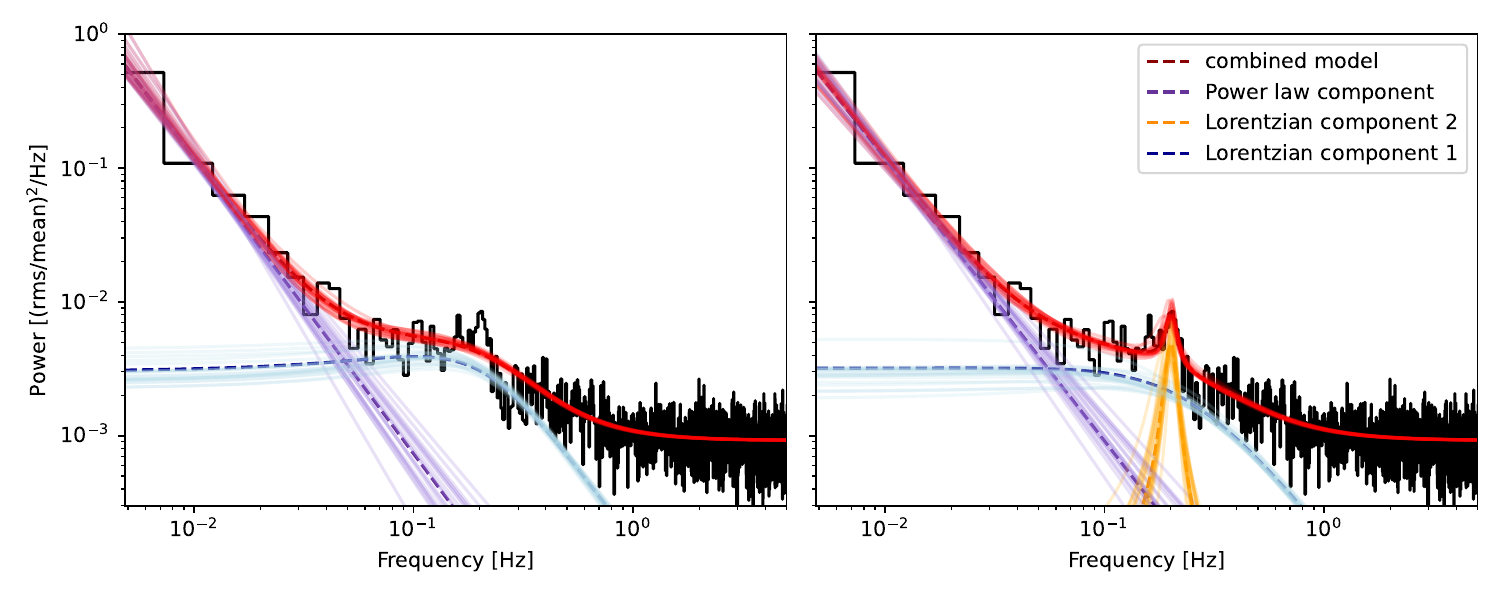}
\caption{Averaged periodogram of \src\ data. In red, we show the combined model for $\mathcal{M}_1$ (left) and $\mathcal{M}_2$ (right), in purple, we show the power-law components, and in blue and orange we show the two Lorentzians (only one for $\mathcal{M}_1$) -- each of which includes a constant white noise component. Dashed lines represent the median of the posterior predictive distribution, and lightly shaded solid lines correspond to twenty individual draws from the posterior predictive distribution, for each of the model components as well as the combined model.}
\label{fig:psd_comp}
\end{figure*}
The typical periodogram observed in accreting XRPs is mainly composed of aperiodical variability (noise) in the form of a low-frequency power-law with a break due to the magnetospheric boundary \citep{Lyubarskii1997}. 
Above the break the power law slope becomes steeper, which is suggestive of a smoother magnetospheric flow \citep{Revnivtsev2009}, typically modeled with a Lorentzian component \citep{Monkkonen2022}.
The average periodogram from \src\ shows a strong noise component at low frequencies, as well as additional excess power in the 0.1-1 Hz band.
In order to assess the significance of the excess power, we fit a model combining a power law for the low-frequency noise component with one Lorentzian for the excess noise (model 1, denoted $\mathcal{M}_1$) or a combination of a power law with two Lorentzian components (model 2, $\mathcal{M}_2$). We denote the Lorentzian components $L_1$ and $L_2$, respectively. 
In both models, we also include a constant $c$ to parameterize the flat noise component at high frequencies (white noise). 
We parameterized the power law in terms of an amplitude, $A_\mathrm{PL}$ and power law index $\alpha$, the Lorentzian with the amplitude $A_\mathrm{Lor}$, the the centroid frequency, $\nu_\mathrm{mid}$ and the quality factor $Q = \nu_\mathrm{mid}/\gamma $, where $\gamma$ is the full width at half maximum.

\begin{figure*}[!t]
\includegraphics[width=0.95\textwidth]{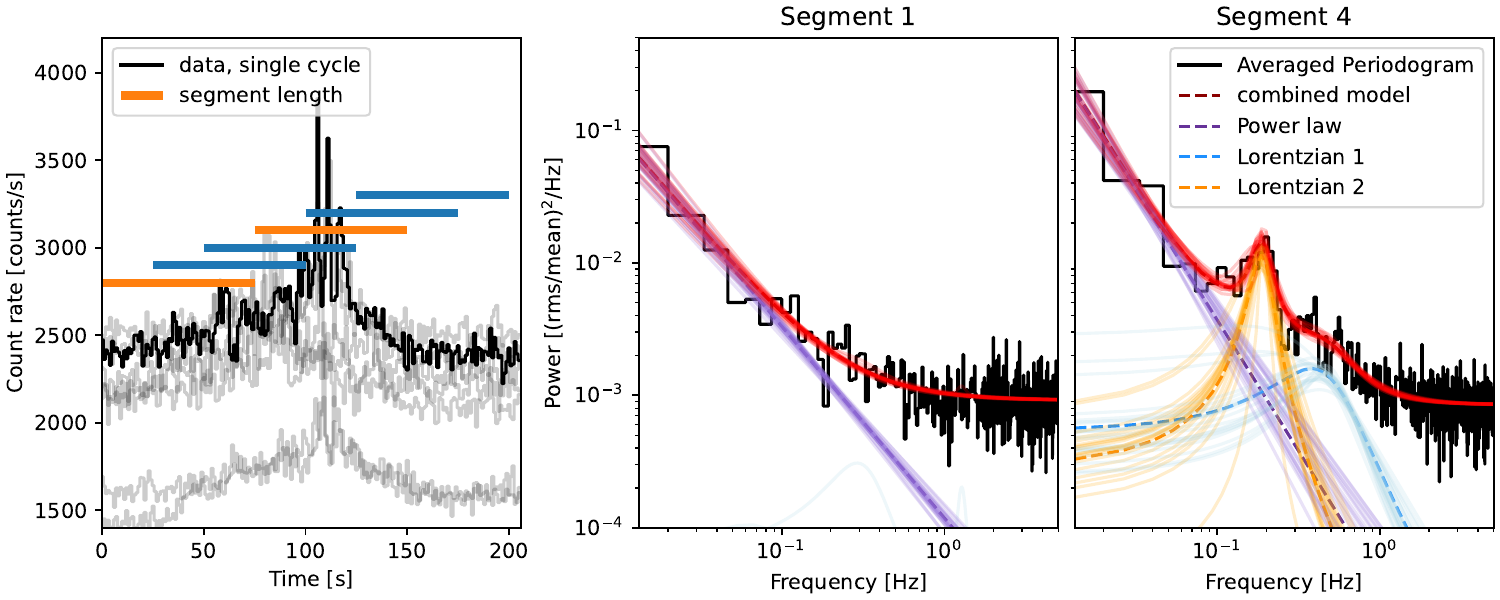}
\caption{Phase-resolved results for \src. \textit{Left:} Aligned light curves (in gray) of the triggers in Table~\ref{tab:GBMtriggers}
-- one highlighted in black for clarity. 
Horizontal bars correspond to the segments used for the phase-resolved analysis. Orange bars represent segments in phase used to produce the right panels.
\textit{Right:} Periodograms of \src\ data at different spin phases, corresponding to segments 1 and 4, as indicated on top. Symbols are as in Fig.~\ref{fig:psd_comp}. The segment 1 panel shows the periodogram resulting from data outside of the pulse profile peak, when the periodogram does not require a second Lorentzian component. The segment 4 panel shows the periodogram obtained from pulse-profile peak-centered data, where a second Lorentzian component is required at $\sim0.2\,$Hz.}
\label{fig:psd_phases}
\end{figure*}

We estimated the Bayes factor $\log(\mathcal{B}_{12})$ to compare the two models and we find $\log(\mathcal{B}_{12}) = - 30.3$, indicating decisive evidence for $\mathcal{M}_2$ (and the presence of a QPO-like component in the periodogram) following the scale of \citet{kass1995} for the interpretation of the Bayes factor.
For $L_2$, we find a centroid frequency of $\nu_{\mathrm{mid}, 2} = 0.199_{-0.004}^{+0.003}$, a quality factor $Q_2 = 16.4_{-5.1}^{+7.4}$, and an amplitude $A_{\mathrm{Lor}, 2} = 0.005_{-0.001}^{+0.002}$, where we report the median of the posterior distribution and the $1\sigma$ credible intervals. 
We report supplementary statistical details of the analysis in Appendix~\ref{sec:A}.
Moreover, given that the strongest variability appears at the peak of the source pulse profile (see Fig.~\ref{fig:230204}), we also obtain periodograms corresponding to different phases of the spin period to test for spin-phase dependence.
To obtain phase-resolved periodograms, we defined six (overlapping) segments lasting 75 s each and covering different phases of the pulse profile, starting 25 s apart (see Fig.~\ref{fig:psd_phases}).
We marked the phase corresponding to the pulse profile peak as 0.5.
For each segment, we modeled the averaged periodogram across all included cycles with the two above-mentioned models, $\mathcal{M}_1$ and $\mathcal{M}_2$.
We find that segments that do not include the time around a phase of $\sim 0.5$ generally favor model $\mathcal{M}_1$, while segments covering the peak of the pulse profile very strongly prefer model $\mathcal{M}_2$ and the presence of a narrow, QPO-like feature, suggesting that the QPO is transient and phase-dependent (see Fig.~\ref{fig:psd_phases} and Appendix~\ref{sec:A} for more details).
Finally, we notice that the limited statistics of the intrinsically transient phenomenon prevents a luminosity-dependent analysis of the QPOs.

Finally, to explore possible spectral variations of the X-ray emission during the observed QPOs episodes, we defined a hardness ratio as the ratio between the photon counts in two energy bands, that is, HR$=20-70\,$keV / $8-20\,$keV. The energy intervals were chosen to ensure comparable statistics in both bands.
We analyzed the HR around the triggers down to a time binning of 0.5 s.
We find that the HR does not show significant variation within the TTE files' durations, sampling a few spin cycles each.

\section{Discussion}
\subsection{QPOs from the Keplerian frequency model}
We discovered the appearance of triple-transient QPOs (ATTO) in \src\ at a frequency of $\sim0.2\,$Hz.
QPOs in the range from mHz to $\sim10\,$Hz have been observed in several XRPs \citep{Reig2008, James2010, Sidoli2016, Nespoli11, Wilson-Hodge18}. 
In some systems, QPOs are persistent, while more often they are transient.
Widely accepted QPO models feature an accretion disk whose inner radius is directly interacting with the NS magnetosphere.
For those models, QPOs are either the result of an accreting inhomogeneity rotating at the magnetospheric radius with the disk Keplerian frequency $\nu_{QPO}=\nu_K(r_m)$ (Keplerian frequency model, or KM, \citealt{vanderKlis89}), or due to variations in the accretion rate at a characteristic frequency equal to the beat frequency between the accretion disk and the NS spin frequency, $\nu_{QPO}=\nu_{beat}=\nu_K(r_m) - \nu_{spin}$, which is the beat frequency model (BFM, \citealt{Alpar1985,Lamb1985}).
The Keplerian frequency at the inner disk radius $r_K$ is equal to
\begin{equation}\label{eq:nu_K}
    \nu_K = \sqrt{\frac{GM_{NS}}{(2\pi)^2r_K^3}},
\end{equation}
where M$_{NS}$ is the NS mass.
From Eq.~\ref{eq:nu_K}, the inner disk radius is r$_K=(GM/(2\pi \nu_K)^2)^{1/3}$.
On the other hand, the magnetospheric radius is $r_m=k\,r_A$, where $r_A$ is the Alfven radius when the energy density of the magnetic field balances the kinetic energy density of the infalling material \citep{DaviesPringle1981}:
\begin{equation}
    r_A = \left(\frac{\mu^4}{2GM_{NS}\dot{M}^2} \right)^{1/7},
\end{equation}
while $k$ is the coupling constant and $\mu=\frac{1}{2}BR_{NS}^3$ is the magnetic dipole moment (with magnetic field strength B and NS radius R$_{\rm NS}$).
Since $\nu_{QPO}\gg\nu_{spin}$, $\nu_{QPO}$ in the BFM is still $\sim\nu_K$. 
Therefore, assuming the KM, we can write ${r_m=r_K}$, from which we find
\begin{equation}\label{eq:mu}
    \mu = 1.9\times10^{31}\,\rm{G\,cm^{3}}\,\it{k}\rm _{0.5}^{-7/4}\,M_{1.4}^{5/6}\,\dot{M}_{-8}^{1/2}\,\left(\frac{\nu_k}{0.2\,\rm Hz}\right)^{-7/6}
\end{equation}
where $M_{1.4}$ is the NS mass in units of $1.4\,M_\odot$, $k_{0.5}$ is the coupling constant in units of 0.5, and $\dot{M}_{-8}$ is the mass accretion rate in units of $10^{-8}M_\odot\,$yr$^{-1}$.
\citet{Salganik2023} finds a luminosity of $\sim3\times10^{37}\,$erg/s which, for an efficiency of $\eta=0.17$, corresponds to a mass accretion rate of $3.1\times10^{-9}\,$M$_\odot\,$yr$^{-1}$. Assuming $k=0.5$, and given that for the keplerian model $\nu_{QPO}\gg\nu_{spin}$ implies $\nu_K = \nu_{QPO}$, then Eq.~\ref{eq:mu} returns a magnetic field of $2.3\times10^{13}\,$G (however, see \citealt{Bozzo09, Bozzo2018} for a discussion about the underlying assumptions in this approach).
\citet{Salganik2023} excluded the presence of a cyclotron line in recent observations of \src\ and used other arguments to estimate a magnetic field of $\sim10^{13}\,$G, in line with our findings.

\subsection{Comparison with EXO 2030+375 and other XRPs}

The QPOs frequency in \src, $\sim0.2\,$Hz, is remarkably similar to that found in another XRP, EXO 2030+375 \citep{Angelini1989}.
Moreover, the newly discovered QPOs share additional similarities with the QPOs observed in other XRPs.
First, similarly to EXO 2030+375 and several other accreting XRPs, the Q factor of the QPOs in \src\ is relatively low when compared to the Q factor of the QPOs observed in other accreting systems \citep{Reig2008}. 
Furthermore, the QPOs discovered in EXO 2030+375 were also transient and detected only around the peak of a bright outburst, at a 1-20 keV luminosity larger than $1.8\times10^{37}\,$erg/s (corrected from the luminosity reported in \citealt{Angelini1989} for a Gaia DR3 distance of $2.4\,$kpc).
Likewise, GBM detections of the QPOs in \src\ start on MJD 59970 at a (1-79 keV) luminosity of about $2.8\times10^{37}\,$erg\,s$^{-1}$ \citep{Salganik2023}, while QPOs are no longer detected after MJD 60000, which is when the source luminosity drops below the same value (see also Appendix~\ref{sec:B} for the non-detection significance).
Although not explicitly reported in previous works, we notice that such behavior is similar to that observed in other XRPs, where QPOs are not detected in the subcritical regime corresponding to the horizontal branch in the hardness-intensity diagram \citep{Reig2008}.
Similar findings are also obtained for the XRP 1A~0535+262 \citep{Finger96_A0535}, where QPOs are observed when the source luminosity is above $6\times10^{36}\,$erg/s (assuming a Gaia DR3 distance of 1.8 kpc), in agreement with the critical luminosity value derived by \citet[see their Fig.~7]{Mushtukov15_crit_lum}, but not when the source luminosity drops below the same value.
However, the opposite is also observed for sources for which QPOs are detected only in the subcritical regime (see, e.g., KS~1947+300, \citealt{James2010}).

It is therefore tempting to associate the appearance of QPOs in \src\ with the onset of the supercritical accretion regime.
Accordingly, we speculate upon the following qualitative scenario.
When the source enters the supercritical accretion regime forming an accretion column at the magnetic poles, the radiation beaming pattern changes from a pencil-beam to a fan-beam pattern \citep{Basko+Sunyaev76, Becker+12}.
Depending on the system geometry, radiation in this regime can be more likely to hit the inner accretion disk (see, e.g., \citealt{Koliopanos+18}). 
The geometry recently proposed for \src\ \citep{Doroshenko2023_LSV} with a magnetic obliquity of $\sim48^{\circ}$ is consistent with this condition (assuming that the accretion disk is perpendicular to the spin axis).
When the radiation hits the disk, if the inner accretion disk viscosity is stochastically unstable in relation to the mass accretion rate and temperature variations (see, e.g., \citealt{Romanova2008}, \citealt{Potter2014}), ``tongues'' of enhanced accretion can form at the disk inner radius $r_K$ and subsequently accrete onto the magnetic poles, thus feeding the observed QPOs.
However, we note that simulations by \citealt{Romanova2008} consider an NS with a relatively low magnetic field typical of millisecond X-ray pulsars. 
It remains to be demonstrated if a phenomenology similar to the tongues revealed by those simulations can also appear in the case of highly magnetized neutron stars.
Besides, QPOs in \src\ are only observed at the peak of the pulse profile. This observational aspect is likely related to the system geometry.
To this aim, we notice that the sharp dip observed in the \src\ pulse profile was interpreted as emission eclipsed by the accretion stream \citep{Usui12, Galloway2001}.
This supports the idea that, in \src, the pulse profile peak represents the emission from the accretion column observed at the closest approach to the observer line of sight.
Under such circumstances, if the QPOs' emission originates at the bottom of the accretion column, when the column becomes occulted by the NS rotation then the QPOs also become invisible to the observer, thus appearing as spin-phase-dependent QPOs. 
Finally, beaming or collimation of the original X-ray emission by the surrounding matter and expelled winds might also influence the appearance of QPOs, although only in pulsating Ultra Luminous X-ray sources are such effects expected to result in highly beamed escaping radiation \citep{Abarca2021,King2023}.
\\

\begin{acknowledgements}
This work has made use of data from the European Space Agency (ESA) mission {\it Gaia} (\url{https://www.cosmos.esa.int/gaia}), processed by the {\it Gaia} Data Processing and Analysis Consortium (DPAC,
\url{https://www.cosmos.esa.int/web/gaia/dpac/consortium}).
O.J.R. gratefully acknowledges NASA funding through contract 80MSFC17M0022. 
\end{acknowledgements}
\bibliographystyle{yahapj}
\bibliography{references}
\filbreak 
\nopagebreak 
\begin{appendix}

\section{Supplementary statistical material}\label{sec:A}

We define a Whittle likelihood \citep{whittle1951hypothesis, chatfield2013analysis} appropriate for averaged periodograms as introduced in \citet{barret2012}. We applied wide, uninformative priors as listed in Table \ref{tab:priors}. We parameterized all parameters that can vary over more than an order of magnitude or that require us to enforce positivity in terms of the logarithm of the parameter. 
Despite the requirement that the amplitude be positive, we note that $\mathcal{M}_1$ is a special case of $\mathcal{M}_2$ if the amplitude of the Lorentzian in the second component is zero. This allows for a relatively straightforward model comparison, and thus we designed $\mathcal{M}_2$ with that model comparison in mind, and a uniform prior on the amplitude itself rather than the logarithm.

We constructed the model using the JAX \textit{python} library \citep{jax2018github}. We sampled the posterior using the Hamiltonian Monte Carlo method, which uses first-order gradient information for efficient exploration of the parameter space, with the No U-Turn Sampler (NUTS; \citealt{homan2014,phan2019,lao2020}) extension as implemented in the Python package \textit{blackjax} \citep{blackjax2020github}. In Figure \ref{fig:psd_comp}, we present the comparison between the two models. 
The second component models a narrow feature at ${\sim}0.2\;\mathrm{Hz}$. 
We note that the posterior distribution for the amplitude, $A_{\mathrm{Lor}, 2} = 0.005_{-0.001}^{+0.002}$, assigns no probability to $A_{\mathrm{Lor}, 2} = 0$, which corresponds to the simpler model $\mathcal{M}_1$. 

We used the nested nature of our models and employed the Savage-Dickey density ratio (SDDR; \citealt{dickey1970}) to estimate the Bayes factor to compare the two models. 
The Bayes factor can be estimated from the marginal posterior density for $p(A_{\mathrm{Lor},2} | D)$ and the prior $p(A_{\mathrm{Lor},2})$. 
We find a Bayes factor of $\log(\mathcal{B}_{12}) = - 30.3$, which strongly favors the $\mathcal{M}_2$ (assuming equal prior probabilities for $\mathcal{M}_1$ and $\mathcal{M}_2$).

There are two notable caveats to this result. Firstly, Bayes factors generally depend strongly on prior assumptions. A sensitivity analysis of the result on the prior for the amplitude of $L_2$ reveals that even dramatically increasing the prior volume by two orders of magnitude only lowers the Bayes factor to $\log(\mathcal{B}_{12}) = -23.4$, which, while lower, would still be interpreted as decisive evidence for $\mathcal{M}_2$. 
The second caveat concerns the use of periodograms for this analysis; the data considered here breaks some of the assumptions made during the statistical analysis of periodograms, most notably stationarity and a purely stochastic process. 
\citet{huebner2021} showed that the presence of non-stationarity in light curves breaks independence assumptions made in the definition of the likelihood, and as a consequence signals may appear more significant than they are in reality. However, we note that the detection here is highly significant (a Bayes factor of $\log(\mathcal{B}_{12}) = -30$ corresponds to odds of ${\sim}10^{14} : 1$), {and we expect it to remain} highly significant even when accounting for the {non-stationarity}. 

Similarly, for the phase-resolved analysis we calculated the Bayes factor $\mathcal{B}_{12}$ using the SDDR for each segment and model choice.
We find (see Fig.~\ref{fig:psd_phases}) that segments not including the time around the phase~${\sim}0.5$ (segments 1, 2, and 6) return Bayes factors~\mbox{favoring} model $\mathcal{M}_1$ ($\log(\mathcal{B}_{12}) = 3.5,\, 0.5,\, 1.4$, respectively), while segments covering the peak of the pulse profile (segments 3, 4, and 5) have Bayes factors that very strongly prefer model $\mathcal{M}_2$ and the presence of a narrow, QPO-like feature at $\sim0.2\,$Hz ($\log(\mathcal{B}_{12}) = -23.9,\, -50.7,\, -42.3$, respectively).

We also notice that individual segments in the phase-resolved analysis necessarily include trends over the duration of the segments, as each segment captures a subset of the overall rise and fall of the pulse period. Trends on timescales longer than the length of a time series will re-distribute power from below the smallest observable frequency into the observable frequency band, an effect called red noise leakage (see, e.g., \citealt{vaughan2003} for a discussion in the context of astronomical X-ray light curves). However, this largely affects the broadband noise properties, that is, the power-law index of the broadband power-law components in our model, not a QPO. The leakage might have a small, subtle effect on QPO detectability owing to the mildly distorted shape of the broadband periodogram against which we detected the QPOs. This effect could be calibrated with extensive simulations. However, we note that the broadband noise components in segments 1, 2, and 6, which largely exclude the phase segments with strong variability, are relatively flat, with a power-law index $\alpha \approx 1.5$. \citet{vaughan2003} suggested that red noise leakage is negligible for similar spectra,
so we expect the effect to be minor compared to the dramatic differences in Bayes factor we find for the different segments.

\begin{table}[!t]
\caption{Priors applied to the two models, $\mathcal{M}_1$ and $\mathcal{M}_2$.}\label{tab:priors}
\centering
\begin{tabular}{lll}
\hline\hline 
Component & Parameter  & Distribution \\ 
\hline
Power law & $\log(A_\mathrm{PL})$ & $\mathcal{N}(-10, 5)$$^b$\\
 & $\alpha$  & $\mathcal{N}(3, 2)$ \\
\hline
Lorentzian 1 & $\log(A_{\mathrm{Lor},1})$ & $\mathcal{N}(-5, 5)$\\
& $\log(\nu_{\mathrm{mid},1})$ & $\mathcal{U}(\log(0.0049), \log(10))$$^c$\\
& $\log(Q_1)$ & $\mathcal{U}(\log(0.01), \log(2))$\\
\hline
Lorentzian 2$^{a}$ & $A_{\mathrm{Lor},2}$ & $\mathcal{U}(0, 1.0)$\\
& $\log(\nu_{\mathrm{mid},2})$ & $\mathcal{U}(\log(0.0049), \log(10)$\\
& $\log(Q_2)$ & $\mathcal{U}(\log(2), \log(50))$\\
\hline
Counting noise & $\log(C)$ & $\mathcal{N}(\log(\mu_\mathrm{n}), \log(\mu_\mathrm{n}/5))$$^d$ \\
\hline
\end{tabular}
\tablefoot{
\tablefoottext{a}{This component is only present in $\mathcal{M}_2$.}
\tablefoottext{b}{$\mathcal{N}(\mu, \sigma^2)$ defines a normal distribution with a mean of $\mu$ and a variance of $\sigma^2$}
\tablefoottext{c}{$\mathcal{U}(a, b)$ defines a uniform distribution with lower bound $a$ and upper bound $b$.}
\tablefoottext{d}{For simplicity, and because the noise in GBM is generally well-behaved, the prior for the white noise component is set using the mean power $\mu_n$ of the last 100 frequencies in the periodogram.}
}
\end{table}

\section{Significance of ATTO non-detection}\label{sec:B}

To demonstrate that the non-detection of QPOs in \src\ below the critical luminosity is not due to GBM sensitivity but that it is a physical effect, we compares the GBM background variability with
the GBM triggering criteria.
To this aim, we sampled TTE hourly data from Jan. 22, 2023 (MJD 59966.759), which is four days before the first GBM trigger from \src (MJD 59970.759) and corresponds to an $\sim50\%$ lower GBM pulsed flux from this source.
This date was chosen because the \textit{Fermi} spacecraft is approximately in the same orbital location and orientation approximately every $\sim48\,$hours \citep{Fitzpatrick2011}; thus the instrument sensitivity is comparable approximately every two days.
At the same time, the sampled date corresponds to the subcritical accretion regime \citep{Salganik2023}.
We verified that source pulsations were observed in the TTE data around MJD 59966.759, although the source did not trigger the instrument.
A GBM trigger occurs when the instrument detects a signal with a significance of $>4.5\sigma$ in at least two NaI detectors.
The typical significance (S/N) of the QPO flares from triggers in Table~\ref{tab:GBMtriggers} is $>10\sigma$ for the detectors with the smallest source angles.
Given the highly variable background in GBM (up to a factor of $\sim2$ over one orbit), we estimate that a QPO flare event similar to those detected on MJD 59970.759 (with SNR${\sim10}\sigma$) would have triggered GBM on MJD 59966 for all possible background rates registered by the detectors with the smallest source angles.
This implies that, although we cannot exclude ATTO occurring at times where GBM had no coverage or had less sensitivity due to particularly noisy events (such as solar flares or incident particles), the strength of the QPO signal, if ever present, has weakened significantly at subcritical luminosity.

\end{appendix}

\end{document}